\documentclass[prl,aps,twocolumn,showpacs,showkeys]{revtex4-2}
\usepackage{amsmath,amssymb,amsthm}

\usepackage{amsfonts,amscd,mathtools,slashed,mathrsfs}

\usepackage{graphicx}% Include figure files
\usepackage{dcolumn}% Align table columns on decimal point
\usepackage{bm}% bold math
\usepackage{marvosym} %piorun sprzecznosciowy
\usepackage{hyperref}
\usepackage{upgreek}
\DeclareMathOperator{\supp}{supp}

\newcommand{\sN}{{\mathbb N}}

\newcommand{\B}{\mathcal{B}}
\newcommand{\C}{\mathcal{C}}

\newcommand{\K}{\mathcal{K}}
\newcommand{\M}{\mathcal{M}}

\newcommand{\cS}{\mathcal{S}}
\newcommand{\U}{\mathcal{U}}

\newcommand{\vc}{\vcentcolon =}             %%  :=  in definition
\newcommand{\cv}{= \vcentcolon}             %%  :=  in definition

\DeclareMathOperator{\vol}{vol}

\usepackage{color}
\usepackage{xcolor}
\usepackage{framed}

\definecolor{shadecolor}{gray}{0.8}
\definecolor{lgray}{gray}{0.5}

\newcounter{mnotecount}[section]
\renewcommand{\themnotecount}{\thesection.\arabic{mnotecount}}
\newcommand{\mnote}[1]%{}
{\protect{\stepcounter{mnotecount}}$^{\mbox{\footnotesize
$%\!\!\!\!\!\!\,
\bullet$\themnotecount}}$ \marginpar{%\color{red}%
\raggedright\tiny\em
$\!\!\!\!\!\!\,\bullet$\themnotecount: #1} }

\definecolor{darkgreen}{rgb}{0,.5,0}

\renewcommand{\ss}{r}
\begin{document}

%\preprint{}

\title{Operational causality in spacetime}

\author{Micha\l\ Eckstein}
\email{michal@eckstein.pl}
\affiliation{Institute  of  Theoretical  Physics  and  Astrophysics,
National  Quantum  Information  Centre,  Faculty  of  Mathematics,  Physics  and  Informatics,
University  of  Gda\'nsk,  Wita  Stwosza  57,  80-308  Gda\'nsk,  Poland}
\affiliation{Copernicus Center for Interdisciplinary Studies, Jagiellonian University, Szczepa\'nska 1/5, 31-011 Krak\'ow, Poland}

\author{Pawe\l\ Horodecki}
\affiliation{International  Centre  for  Theory  of  Quantum  Technologies,  University  of  Gda\'nsk,  Wita  Stwosza  63,  80-308  Gda\'nsk,  Poland}
\affiliation{Faculty of Applied Physics and Mathematics, National Quantum Information Centre,
Gda\'nsk University of Technology, Gabriela Narutowicza 11/12, 80-233 Gda\'nsk, Poland}

\author{Tomasz Miller}
\affiliation{Copernicus Center for Interdisciplinary Studies, Jagiellonian University,
Szczepa\'nska 1/5, 31-011 Krak\'ow, Poland}

\author{Ryszard Horodecki}
\affiliation{Institute  of  Theoretical  Physics  and  Astrophysics,\,
National  Quantum  Information  Centre,  Faculty  of  Mathematics,  Physics  and  Informatics,
University  of  Gda\'nsk,  Wita  Stwosza  57,  80-308  Gda\'nsk,  Poland}

\date{\today}

\begin{abstract}
The no-signalling principle preventing superluminal communication is a limiting paradigm for physical theories. Within the information-theoretic framework it is commonly understood in terms of admissible correlations in composite systems. Here we unveil its complementary incarnation --- the `dynamical no-signalling principle' ---, which forbids superluminal signalling via measurements on simple physical objects (e.g. particles) evolving in time. We show that it imposes strong constraints on admissible models of dynamics. The posited principle is universal --- it can be applied to any theory (classical, quantum or post-quantum) with well-defined rules of calculating detection statistics in spacetime.  As an immediate application we show how one could exploit the Schr\"odinger equation to establish a fully operational superluminal protocol in the Minkowski spacetime. This example illustrates how the principle can be used to identify the limits of applicability of a given model of quantum or post-quantum dynamics. 
% The consequent logical paradox is avoided by recognising that the characteristic length and time scales of causality violation lie beyond the domain of model's applicability.
\end{abstract}

                             % Classification Scheme.s
\maketitle

%%%%%%%%%%%
%\section{Introduction}
%%%%%%%%%%
{\it Introduction}.--- 
The problem of causality in quantum theory, ignited by the famous Einstein--Bohr debate \cite{EPR,BohrEPR}, has long been a controversial topic. 
It took a few decades to realise that although quantum correlations are stronger than the ones available classically, they do not allow for superluminal transfer of any information \cite{NoSignalling,NoCloning}. The latter demand, known as the \emph{no-signalling principle} is now recognised as an essential feature of any physical theory. It prevents the logical inconsistencies that might emerge from the incompatibility of correlations between space-like separated events with the causal structure of spacetime. While met in both classical and quantum physics, it turned out to leave room for post-quantum theories \cite{PR_box,QuantumCausality,PawelRaviCausality,Coecke2,DAriano,InformationCausality,BruknerX,Hardy2005}.

Yet, there exists a second face of no-signalling connected with the inherent dynamics of simple physical systems, such as the time-evolution of a single particle. In the domain of classical physics, the existence of tachyons (when allowed to interact with the ordinary matter) would readily imply the possibility of superluminal communication. However, quantum particles do not have a definite position in space when propagating \emph{per se}, so the  `classical' protocols of information transfer do not automatically apply. Furthermore, the quantum measurement
effectuates a dramatic change on particle's dynamics.
In consequence, although several formal results \cite{Hegerfeldt1,Hegerfeldt2,Hegerfeldt1985,HegerfeldtFermi,WSWSG11,PRA2017} suggested that quantum wave packets can propagate superluminally, it was unclear whether it implies operational faster-than-light communication (cf. for instance \cite{HegerfeldtFermi} vs \cite{Yngvason} or a more recent work \cite{Pavsic2018}).

\begin{figure}[h]
\begin{center}
\resizebox{0.48\textwidth}{!}{\includegraphics{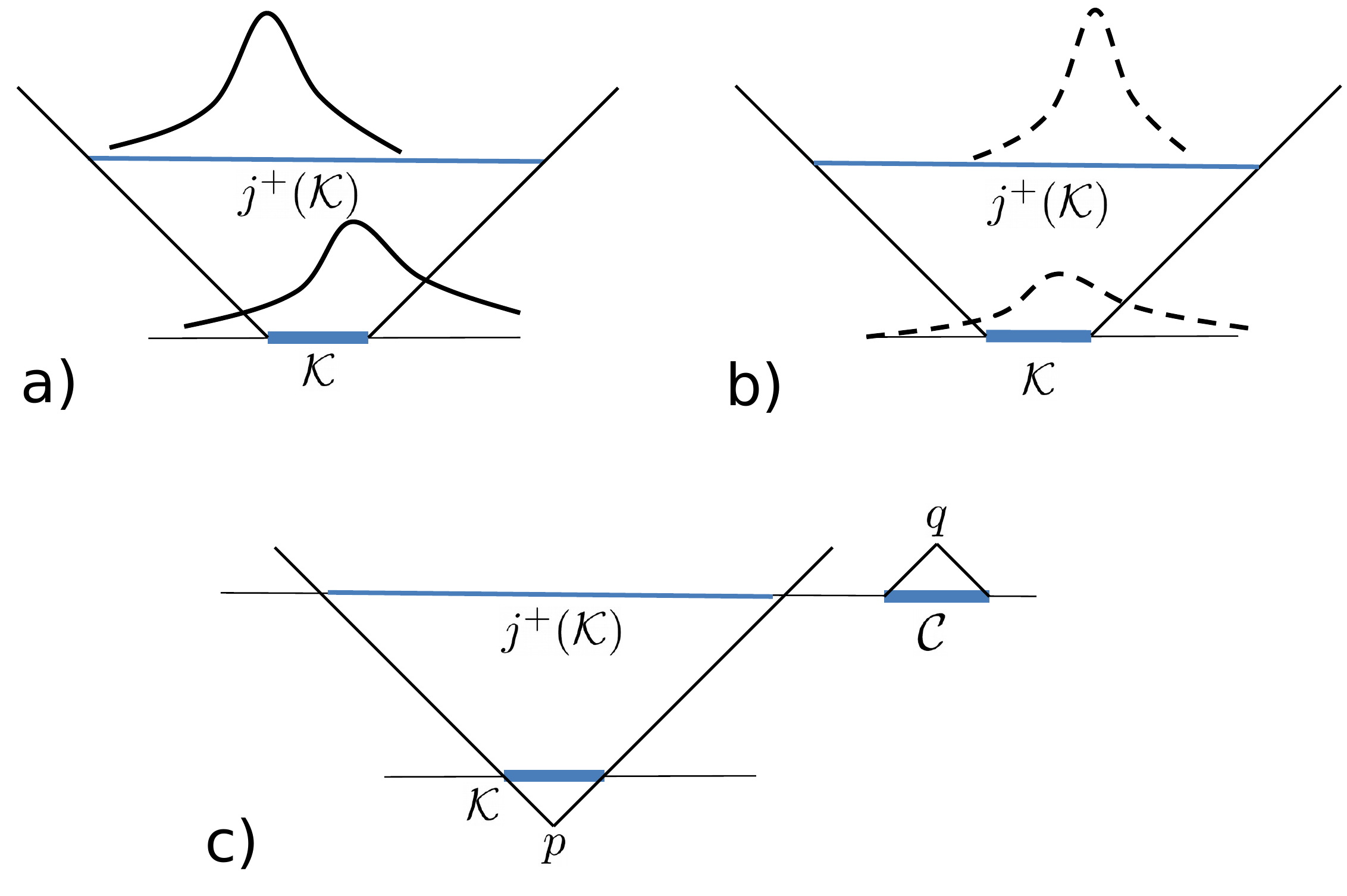}}
\caption{\label{Fig:main} Summary of the main result:
\textbf{a)} In classical physics, a causal propagation of a probability measure (modelling a spatially distributed physical quantity) must satisfy an optimal-transport-theoretic condition \eqref{axiom0}, which encodes the demand that infinitesimal portions
of probability cannot move faster than light or --- equivalently --- that the probability `mass' initially 
contained within an compact set $\K$ must stay in its causal future $j^+(\K)$.
\textbf{b)} The propagation of a quantum wave packet gives rise to a certain ``potential detection 
statistics'', which does not exist objectively until an actual measurement. 
Nevertheless, the present result says --- surprisingly --- that the causal structure of spacetime
coerces the same optimal-transport-theoretic condition \eqref{axiom0} on the statistics of \emph{unperformed} (sic!) experiments. 
\textbf{c)} The violation of condition \eqref{axiom0} leads to an operational protocol of 
faster-than-light signalling. Namely, the free decision ``to perform or not to perform''
a measurement in the set $\K$ taken at a spacetime point $p$ results in
 a change of the potential detection statistics in some other set $\C$ outside of 
the future of both the set $\K$ and the event $p$. Since this statistics can be collected at a later 
event $q$, this leads to statistical signalling between the spacelike-separated events $p$ and $q$.
 }
\end{center}
\end{figure}

In this article we formulate the \emph{dynamical no-signalling principle}, which says that one must not be able to exploit the inherent dynamics of any physical phenomenon for superluminal signalling. From this standpoint we identify the operational constraints on both dynamics and measurement schemes. The adopted formalism of measures on spacetime is very general and can be applied in any theory --- classical, quantum or post-quantum. We show that, surprisingly, any conceivable dynamics must abide by a strong classical constraint related to the evolution of point-like particle's statistics (see Fig. \ref{Fig:main}). As a concrete application, we resolve the controversy around the purported (a)causality of wave packet evolution. This is done by firstly demonstrating that the Schr\"odinger equation together with a local detection can lead to a logical paradox and, secondly, by recognising the pertinence of the physical time and length scales of causality violation. The latter provide ultimate bounds on the domain of applicability of a given model of dynamics.

\medskip
{\it Measurements and communication in spacetime}.--- 
Our starting point is a spacetime $\M$ consisting of events. The latter are associated with classical (i.e. objective) information and form the basic elements of any information processing protocol (cf. \cite{Pandora}). The spacetime has an inbuilt causal structure, that is a partial order relation $\preceq$ specifying which events might influence one another. If such an influence between two events is precluded, we say they are \emph{spacelike-separated}. Furthermore, a spacetime must admit (at least locally) a splitting $\M \cong I \times \cS$ into time range $I$ and space $\cS$. Such a splitting need not be unique and typically depends upon the choice of the reference frame. 

It is standard to accept that the causal structure of spacetime is defined by the constant speed of light in vacuum, but other options are also conceivable (cf., for instance, \cite{GlobalFoliation}). In any case, we shall say a dynamical law is \emph{acausal} or \emph{superluminal} if it implies influences between spacelike-separated events. The incompatibility of dynamics with the assumed casual structure typically leads to logical paradoxes, which spoil the model's consistency (see Fig. \ref{paradox} for an explicit example).

The simplest two-party communication protocol entails two definite events: signal sending $p$ and signal reception $q$. The sender (Alice) is active --- she can freely choose the bit she desires to communicate. On the other hand, the receiver (Bob) passively gathers the incoming information. The communication is effective if Alice's free choice changes Bob's detection statistics registered during the event $q$. Clearly, if the change in Bob's statistics is tiny then Alice should send multiple copies of her signal. The communication is called superluminal if $q$ is not in the causal future of $p$.

A probability measure $\mu$ supported on $\cS_t := \{t\} \times \cS$ is a natural mathematical object tailored to model the statistics of a basic binary measurement at a given time $t \in I$. Indeed, if a(n array of) detector(s) is located in a compact region of space $\K$, $\mu$ yields a concrete number $\mu(\K) \in [0,1]$. The compactness of $\K$ reflects the demand of the locality of the measuring device (cf. \cite{Haag,QIandGR}).

Let us stress that the ``detection'' is purely operational and signifies a detector click --- which constitutes an objective bit of information --- registered in the region $\K$. The interpretation is secondary and depends on the theory. For instance, if the signal had been carried by a classical-like particle the click would mean that the detection was an actual event located somewhere in $\K$. Had it been a quantum particle instead, the only admissible conclusion would be that the particle is not outside of $\K$ at moment $t$ with certainty.

The time-evolution of a probability measure on a spacetime $\M$ is defined \cite{PRA2017,Miller17a} as a map $t \mapsto \mu_t$ such that $\supp \mu_t \subset \cS_t$. It models the time-evolution of \emph{purely potential} detection statistics associated with the dynamics of a simple physical system, such as a propagating quantum particle. The number $\mu_t(\K)$ answers the question: What is the probability of signal detection \emph{if} a detector covering a region $\K$ is switched on at a time moment $t$. The basic signalling task requires two local operations, so only the initial and final measures --- $\mu \vc \mu_s$ and $\nu \vc \mu_t$, respectively --- are relevant. 

Consider now an actual measurement checking whether the signal at an initial time $s$ is within the compact set $\K$. The impact of the measurement on the final measure at some later time $t$ is taken into account with the help of conditional measures
\begin{align*}
\nu( \, \cdot \, | m_{\K}), \ m_{\K} \in \{ 0, 1 \}, 
\end{align*}
where $m_{\K} = 1$ ($m_{\K} = 0$) corresponds to the situation when the measurement has been (has not been) performed at time $s$. The statistics of the measurement $\{ P(\ss|m_{\K}) \}$ with the possible results $\ss \in \{+,-,\emptyset\}$ (corresponding to ``signal detected'', ``signal not detected'', ``not applicable'', respectively) satisfies the rules:
\begin{align}
\label{consistency1}
\begin{array}{lll} 
P(+|1) = \mu(\K), \ \ & P(-|1) = 1 - \mu(\K), \ \ & P(\emptyset|1) = 0, 
\\
P(+|0) = 0, \ \ & P(-|0) = 0, \ \ & P(\emptyset|0) = 1.
\end{array}
\end{align}
Another consistency condition:
\begin{align}
\label{consistency2}
\nu(\, \cdot \,|0) = \nu
\end{align}
must hold, because the absence of the measurement does not disturb the dynamics. 

\medskip
{\it Causal evolution of statistics}.--- 
A classical particle is bound to travel along a future-directed causal curve. In other words, its propagation marks a one-parameter family of events $p_t$ for $t \in I$, with $p_s \preceq p_t$ for all $s \leq t$. Because classical particles carry objective information, any ``tachyon'' violating the causal order $\preceq$ transfers information to a forbidden region of spacetime, eventually resulting in the logical inconsistency of the model.

This classical picture has been formally extended in \cite{AHP2017,PRA2017} to the measure-theoretic setting (see Fig. \ref{Fig:main} a)):\\

\noindent\emph{Causal Evolution (CE) condition}.- The inequality
\begin{align}
\label{axiom0}
\mu_s(\K) \leq \mu_t(j^+(\K)),
\end{align}
with $j^+(\K) \vc J^+(\K) \cap \cS_t$, must hold for all compact $\K \subset \supp \mu_s$ and for all $s \leq t$. \\

The condition CE is a relativistic invariant, i.e., it does not depend upon the adopted splitting $\M \cong I \times \cS$ \cite{Miller17a}. It encodes, via the theory of optimal transport (cf. \cite{AHP2017,Miller17a}), the following classical intuition:\\

\emph{Each infinitesimal part of the probability distribution must travel along a future-directed causal curve}.\\

Such a demand has a clear justification when $\mu$ models the statistical distribution of an ensemble of classical particles, e.g. dust or fluid. In this context, condition \eqref{axiom0} grasps the demand that none of the elements constituting the considered physical medium can propagate superluminally.

However, it seems unlikely that CE might be the proper incarnation of the no-signalling principle beyond the classical realm. Firstly, the mere fact of the existence (and of dynamical emergence) of statistical dependencies between spacelike separated parts of a physical system, does not necessarily imply the possibility of superluminal information transfer, neither in quantum mechanics, nor in a ``post-quantum'' theory with stronger correlations \cite{PR_box}. Secondly, CE treats the detection statistics \emph{as if they objectively existed}, whereas in quantum mechanics what evolves is purely potential --- `non-existing' --- statistics. Furthermore, CE makes no reference to the actual detection process, which, in quantum theory, does change the signal's dynamics. Finally, there is no straightforward connection between the violation of CE for some set $\K$ and a physical signalling process, which involves two definite events --- i.e., points of $\M$ rather than sets.

Curiously enough, it turns out that the violation of CE, when complemented by a minimalistic detection scheme, \emph{always} leads to operational superluminal communication. This surprising conclusion holds independently on whether the detection statistics actually exists before the measurement or not.

\medskip
{\it Dynamical no-signalling}.---
Many of the physical detection events involve the demolition of the signal carrier --- most notably, the photon absorption in a silicon detector. After such an event, the recorded information ``click'' or ``no-click'' is objective, hence the following demand is indispensable:\\

{\it Axiom 1 (A1)}.- If the signal has been detected ($\ss=+$) at time $s$ in the region $\K$ ($m_{\K}=1$), then it must be present with certainty in that region's future $j^+(\K)$ for any later time $t$:
\begin{align}
\label{axiom}
\nu(j^+(\K)|+,1)=1.
\end{align}

Let us stress that the adopted detection scheme includes the von Neumann measurement, but is not limited to it. It can be applied equally well in a non-demolition scenario (cf., for instance, \cite{Haroche2007}), in which case  after the detection the particle (or whatever signal carrier one considers) might undergo an entirely different dynamics than before.
 
The intuition that the detection process in a region $\K$ must not affect the (potential) statistics outside of $J^+(\K)$ is formalised with the help of conditional measures:\\

\noindent{\it Dynamical no-signalling condition (NS)}.- For any compact $\C \subset \cS_t \setminus j^+(\K)$,
\begin{align}
\label{ns}
\nu( \C \, | \, 1) = \nu( \C \, | \, 0).
\end{align}

Strikingly, it turns out that the intuitive Axiom 1 together with condition NS jointly imply that the dynamics of measures must obey the strong classical-like constraint CE. Equivalently,

{\it Proposition 1.- Under the assumption of A1, the violation of CE entails the violation of NS.}

The somewhat technical proofs of Proposition 1, and of Theorem 2 below, are included in the Supplemental Material.

Condition NS, while intuitive, entails information flow from \emph{region $\K$ to region $\C$}, whereas operational signalling requires definite events. 
Nevertheless, it turns out that the violation of NS can always be exploited for a superluminal communication between two strictly local agents.     

{\it Theorem 2.- If NS is violated, the set $\C$ for which (\ref{ns}) does not hold can always be chosen so that there exist spacetime points $q, p_1, \ldots, p_k$ such that
\begin{align*}
\K \subset \bigcup\limits_{i = 1}^k J^+(p_i), \ \ \C \subset J^-(q), \ \ \textnormal{and} \ \ p_i \not\preceq q, \ i=1,\ldots,k.
\end{align*}
}

\begin{figure}[h]
\begin{center}
\resizebox{0.43\textwidth}{!}{\includegraphics{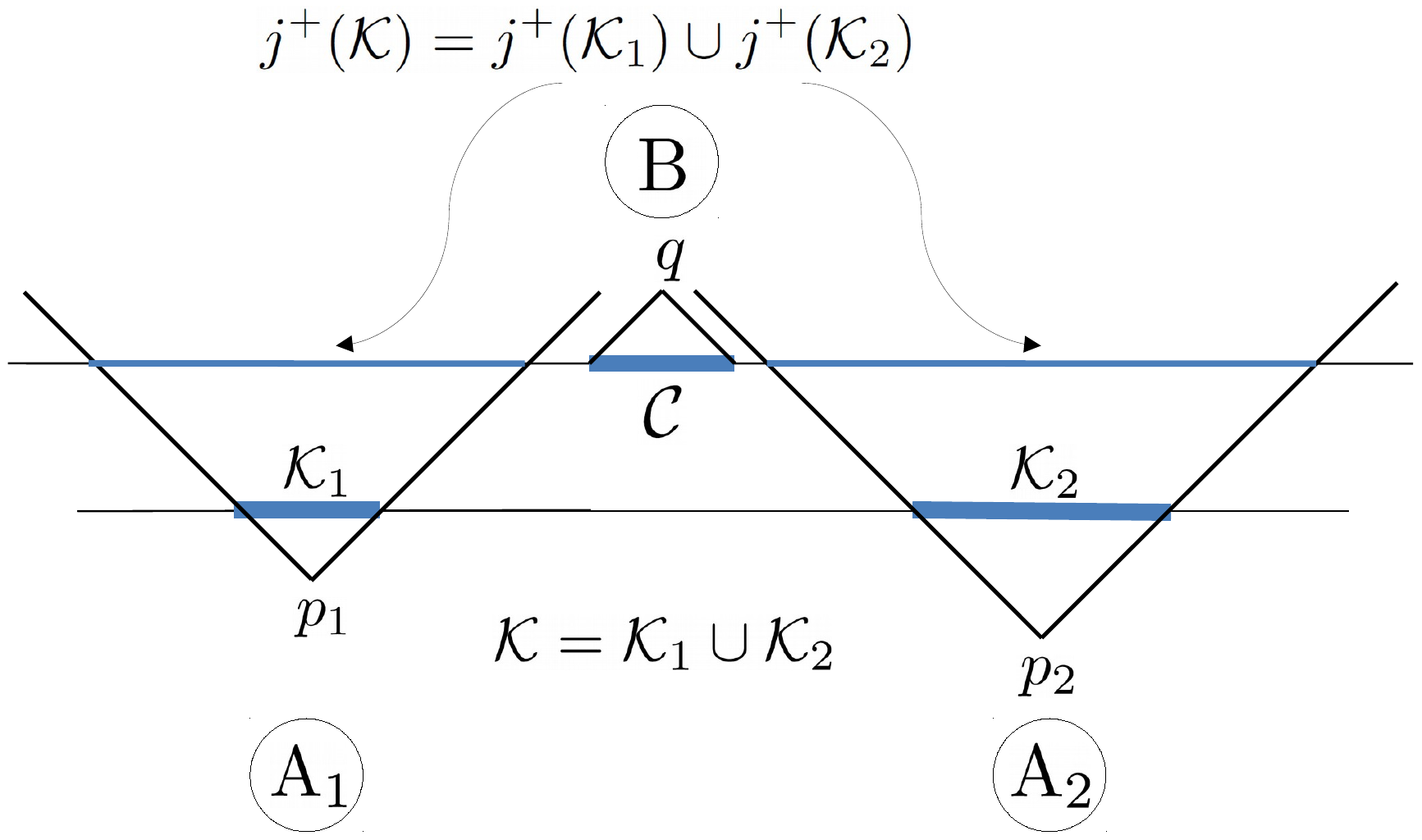}}
\caption{\label{fig1} Diagram illustrating an instance of Theorem 2 with two senders: a passive one ``Alice 1'' and an active one ``Alice 2''.}
\end{center}
\end{figure}

The essence of Theorem 2 is illustrated in Fig. \ref{Fig:main} c). In order to send a superluminal signal we shall first fill both regions $\K$ and $\C$ with detectors. The violation of NS implies that the measurement effectuated by devices in $\K$ changes the detection probability in $\C$. However, to actually execute the (statistical) superluminal signalling we would firstly need to orchestrate the measurement in $\K$ and, secondly, gather the statistical information from $\C$. This amounts to the existence of sending event(s) $p_1, \ldots, p_k$ and a readout event $q$, such that $p_i \npreceq q$ for all $i$.

In the simplest scenario ($k=1$) there is a single sending event and so statistical signalling from $p_1$ to the spacelike-separated $q$ is straightforward. Theorem 2 says that the set $\C$ can always be chosen in such a way that the readout is performed at a single event $q$. This is exactly the situation depicted in Fig. \ref{Fig:main} c).

On the other hand, the set $\K$ might have a more complicated shape (see Supplemental Material for an example), in which case $k > 1$ sending events are needed. For concreteness let us assume that $k =2$, as in Fig. \ref{fig1}. Observe first that a measurement performed just in one of the regions $\K_1$ or $\K_2$ does not lead to the violation of NS, for if it would do so a single sending event would suffice. Consequently, we can assume that one of the senders, say ``Alice 1'' is passive and has an always-on detector ($m_{\K_1} = 1$). The active sender ``Alice 2'' decides whether to perform a measurement or not ($m_{\K_2} = 1$ or $0$), which fixes the value of the communicated bit because $m_{\K} = m_{\K_1} \cdot m_{\K_2} = m_{\K_2}$. The value of this bit can be statistically inferred by Bob from the difference between $\nu( \C \, | \, m_\K = 1)$ and $\nu( \C \, | \, m_\K = 0)$. The generalisation to $k > 2$ senders is straightforward. 

{\it Conclusion.- A violation of NS always has operational consequences --- it enables a protocol suitable for operational superluminal communication.}

We have presented the constraint on free evolution of potential detection statistics and the admissible change of measure upon the positive result of the detection by Alice. For completeness, let us now discuss the effect of a negative result of the detection.

\medskip
\medskip
{\it Complementary axiom and the triad of interrelated conditions}.---
Just as Axiom 1 puts constraints on the possible evolution of the measure $\mu( \, \cdot \, |+,1)$, i.e., the measure conditioned upon the \emph{positive} result of the detection measurement in $\K$, the following condition deals with the measure $\mu( \, \cdot \, |-,1)$, i.e., conditioned upon the \emph{negative} result.\\

{\it Axiom 2 (A2)} .- If the signal has \emph{not} been detected at time $s$ within $\K$, then outside of $J^+(\K)$ the evolution of $\mu(\, \cdot \, | \, - , 1)$ proceeds with no modification other than re-normalisation:
\begin{align}
\label{c2_1}
\nu( \C \, | \, - , 1) = \frac{\nu(\C)}{1 - \mu(\K)},
\end{align}
for any compact $\C \subset \cS_t \setminus j^+(\K)$.\\

This axiom can be intuitively justified as follows: Immediately after the measurement, the result of which was positive, we must have $\mu(\K | +,1) = 1$ and thus $\mu(\, \cdot \, |+,1)$ is zero outside of $\K$. Moreover, irrespectively of the measurement's result, the statistics of the potential measurements outside of $\K$ must remain unchanged so as not to allow for instantaneous signalling, i.e., $\mu(\K'|1) = \mu(\K'\,|\,0)$ for every $\K' \subset \cS_s \setminus \K$. Altogether, using subsequently \eqref{consistency2}, \eqref{ns} and \eqref{consistency1}, one gets
\begin{align*}
\mu(\K') & = \mu(\K'|0) = \mu(\K'|1)
%\\
%& = \tilde{\mu}(\K',+|1) + \tilde{\mu}(\K',-|1)
\\
& = \underbrace{\mu(\K'|+,1)}_{= \, 0}P(+|1) + \mu(\K' |-,1)P(-|1)
\\
& = \mu(\K' |-,1)(1 - \mu(\K)).
\end{align*}
We see that $\mu(\K' |-,1) = \mu(\K')/(1 - \mu(\K))$ for any $\K'$ disjoint with $\K$. Since the evolution outside of $J^+(\K)$ should not be altered, this formula `time-evolves' into \eqref{c2_1}.

The three conditions A1, A2, NS, together with the overarching constraint CE, turn out to enjoy an intimate logical interplay captured by the following result: 

{\it Theorem 3. If any two conditions from the set $\{ \textrm{NS}, \textrm{A1}, \textrm{A2} \}$ hold true, then the third one and CE hold true as well.}
\begin{figure}[h]
\begin{center}
\resizebox{0.25\textwidth}{!}{\includegraphics{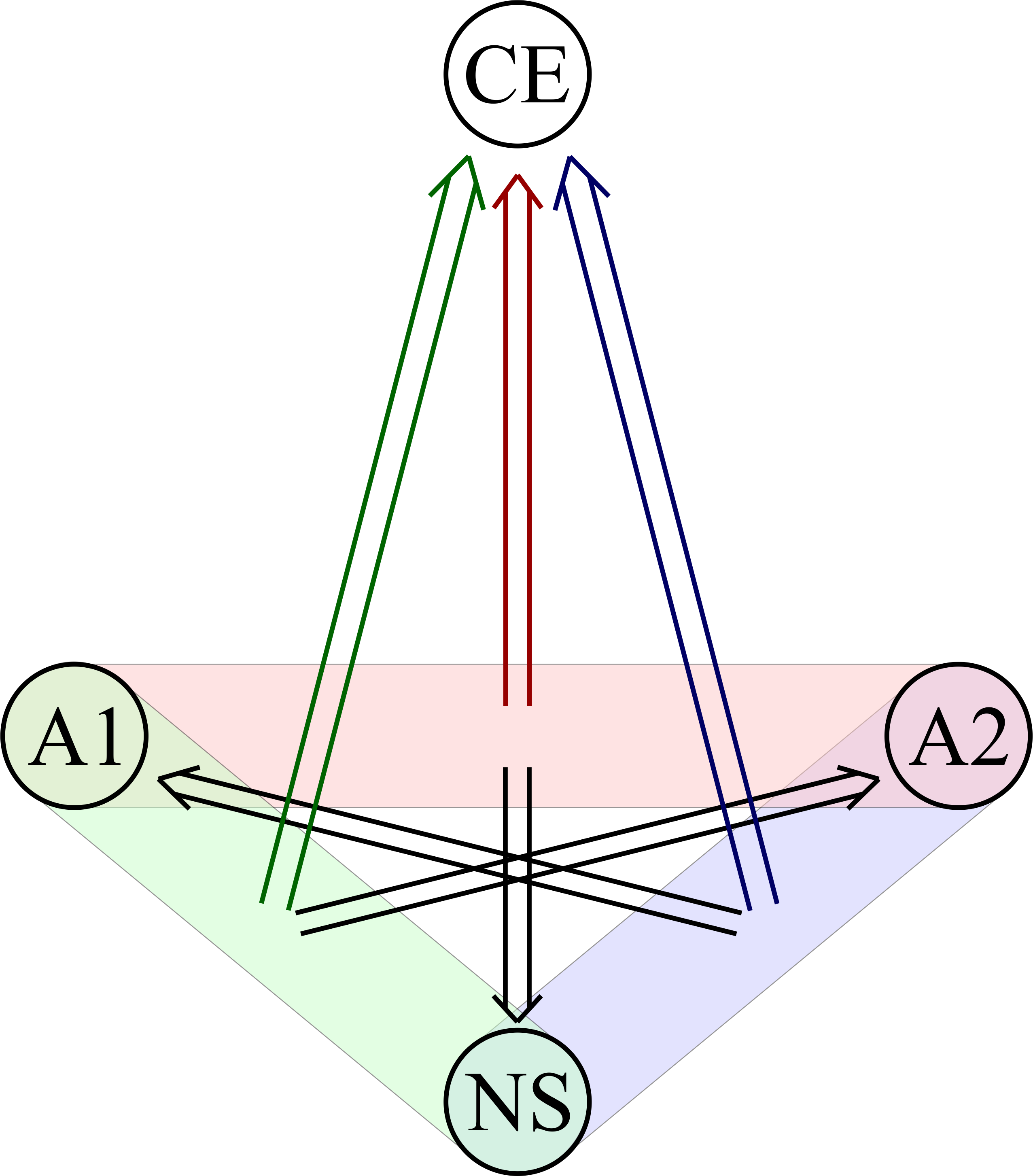}}
\caption{Diagram illustrating Theorem 3.}
\end{center}
\end{figure}

In fact, Theorem 3 exhausts the logical dependencies between the four conditions considered. More rigorously speaking: Any combination of the true/false values assigned to the conditions NS, A1, A2, CE which is not deemed impossible by Theorem 3, can be realised with suitably defined measures $\mu$, $\nu$ and $\nu(\, \cdot \, | \, \pm , 1)$. For more details, as well as for the proof of Theorem 3, the Reader is invited to consult the Supplemental Material.

\medskip
{\it An illustration: logical paradoxes from quantum wave dynamics}.---
Consider two observers, active Alice (A) and passive Bob (B), in the Minkowski spacetime. They have at their disposal a quantum wave packet following a unitary evolution driven by a Hamiltonian $\hat{H}$. Bob has a(n always-on) binary detector covering a region $\C$. Now, consider two situations:  

1) Suppose first that Alice can emit (prepare) a wave packet $\psi_0$ compactly supported within the region $\K$, spacelike-separated with $\C$. She decides (event $p$) to emit ($m_\K = 1$) or not to emit it ($m_\K = 0$). According to Hegerfeldt's theorem \cite{Hegerfeldt1,Hegerfeldt2} if the Hamiltonian $\hat{H}$ is bounded from below, the initial wave packet $\psi_0$ immediately becomes spread over the entire space. Consequently, A1 is violated. Moreover, CE is violated, because $\mu(\K) = 1$ and $\nu(j^+(\K)) < 1$. Finally, NS is also violated because $\nu(\C \, | \, 0) = 0$ (for the wave packet has not been emitted in the first place), but $\nu(\C \, | \, 1) > 0$.

2) Suppose now that Alice cannot prepare localised states, but she can collapse (event $m_\K = 1$) a preexisting quantum wave packet $\psi_0$ supported over entire space $\cS_s$. Alice registers a click ($m_\K = 1, r = +$) with probability $\mu(\K) < 1$, in which case $\nu(\C \, | +, \, 1) = 0$, because the particle is no more. In such a situation, Axiom 1 is readily verified. However, if Alice's detector fails to click, although it was on ($m_\K = 1, r = -$), Bob has a non-vanishing probability of detecting it, $\nu(\C \, | -, \, 1) > 0$. If the dynamics is such that $\nu(\C \, | \, 0) = (1-\mu(\K)) \nu(\C \, | -, \, 1) = \nu(\C \, | \, 1)$ (which is exactly Axiom 2), then there is no superluminal signalling, because Bob cannot statistically distinguish the situations ($m_\K = 1$) and ($m_\K = 0$). However, if the dynamics violates CE, then --- by Proposition 1 --- NS is also violated. Consequently, Axiom 2 fails as well.

In either case, if Alice can statistically signal to Bob, she might fall into a logical paradox. Indeed, suppose that Alice is free-falling and Bob travels away from Alice with a constant relativistic velocity (see Fig. \ref{paradox}). Alice triggers the superluminal protocol (event $p$, $m_\K = 1$). If the bit $m_\K$ is successfully communicated to the region $\C$ spacelike-separated with $\K$, then Bob (automatically) triggers his protocol (event $p'$, $m_{\K'} = 1$). In consequence, the bit $m_{\K'}$ is in turn statistically communicated to the region $\C'$ spacelike-separated with $\K'$. If the signalling succeeds, then Alice receives the bit $m_{\K'}$ (event $q'$). But because $q' \preceq p$, such an event might, for instance, destroy Alice's laboratory, thus extorting $m_\K = 0$.

Note that this scenario is covariant in the sense that if we interchange the roles of Alice and Bob, Bob can (statistically) effectuate a causal loop $p' \rightsquigarrow q$ with the passive help of Alice.

\begin{figure}[h]
\begin{center}
\resizebox{0.49\textwidth}{!}{\includegraphics{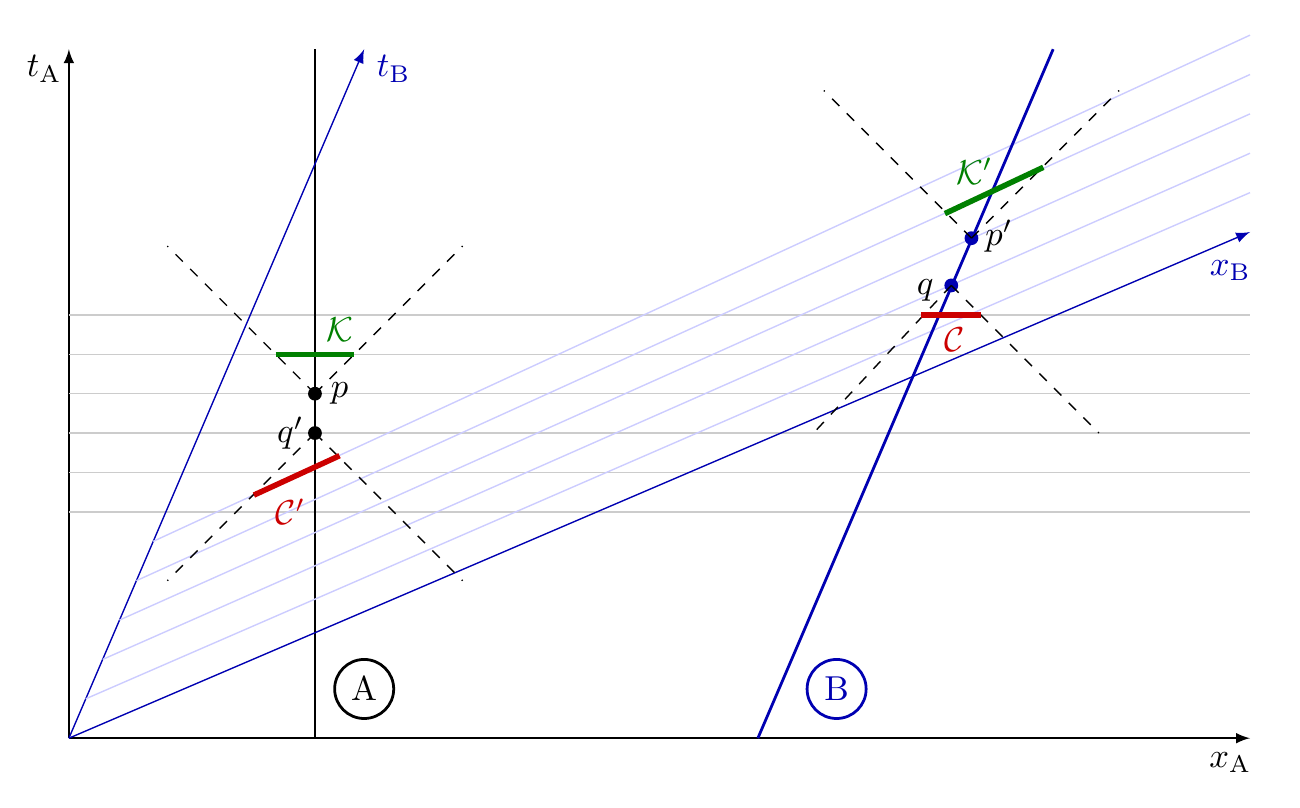}}
 \caption{\label{paradox}Logical paradoxes from superluminal wave packet spreading in the Minkowski spacetime.}
\end{center}
\end{figure}

The problem(s) with situation 1) are well known and are usually bypassed by a QFT theorem on the non-existence of strictly localised states \cite{ReehSchlieder}. On the other hand, situation 2) (analysed in \cite{Hegerfeldt1985} and \cite{AHP2017}, but without the key role of active measurements) harmonises with the view that whereas quantum states are inherently non-local, the operations are local \cite{Haag,QIandGR}. In this context, it is rather surprising to learn \cite{AHP2017} that the unitary evolution under the standard free Hamiltonian $\hat{H} = \hat{p}^2 /2m$ \emph{does violate} the condition CE. The fact that the \emph{relativistic} Hamiltonian $\sqrt{\hat{p}^2 + m^2}$ also implies the violation of CE \cite{AHP2017}  (cf. also \cite{Hegerfeldt1985}) is even more abstruse. In view of the generality of Hegerfeldt's result, one might expect that the violation of CE by quantum wave dynamics is generic. The notable exceptions are e.g. Dirac and photon wave function (Maxwell) equations, which guarantee CE for any (even compactly supported) initial state \cite{AHP2017}. Does it mean that all other quantum wave dynamics, including the standard Schr\"odinger equation, should be discarded as entailing logical paradoxes? 

The appeasement comes from the consideration of the \emph{characteristic scales} of causality violation. Indeed, the explicit account of the spacetime aspect naturally establishes the time-scale $t-s$ of signalling time-lapse, the size $\vol \K$ of signalling devices and the ``capacity of the superluminal channel'' $\mathrm{max} \{0,\nu(\C|0) - \nu(\C|1)\}$.

For example (see \cite{PRA2017}), one can safely use the relativistic Hamiltonian $\sqrt{\hat{p}^2 + m^2}$ for modelling the dynamics of a single quantum particle. Indeed, the causality violation effects in this model are transient and restricted to a region of the size of the particle's Compton wavelength --- a regime in which the quantum nature of the vacuum can no longer be neglected. On the other hand, the superluminal spreading of a Gaussian wave packet driven by the non-relativistic Hamiltonian $\hat{p}^2/(2m)$ induces immediate and persistent causality violation effects for $\K = [-\ell,\ell]^3$, with $\ell > \tfrac{c m \lambda^2}{t \hbar^2} \left(m \lambda^2 + \sqrt{m^2 \lambda^4 + t^2 \hbar^2} \right)$, where $\lambda$ is the width of the initial wave packet. Nevertheless, we observe that the minimal scale of causality violation in this model grows with $m$ and $\lambda$, even in the limit of an infinite time-lapse. Consequently, for instance, at the scales characteristic to BEC ($m \sim 10^{-26} \,\mathrm{kg}, \lambda \sim 1 \,\mathrm{\upmu m}$), the superluminal spreading would manifest itself only for $\ell \gtrsim 30 \,\mathrm{km}$.

\medskip
{\it Discussion}.--- We have shown that the embedding of information processing protocols within a spacetime unravels the ``dynamical no-signalling principle'', which is complementary to the one exploiting correlations. When a minimalistic assumption about the measurement scheme is adopted, the principle coerces a strong constraint on the dynamics of detection statistics. Although the latter embodies the concept of transport along causal curves taken from classical physics, it applies to \emph{any} model within quantum theory or even beyond.

When applied to quantum wave dynamics the unveiled principle leads to an arresting consequence: \emph{The Schr\"odinger equation facilitates operational superluminal signalling via local detection.}  Our finding reinforces and extends the earlier claims around   Hegerfeldt's theorem (cf., for instance, \cite{Hegerfeldt1998_ann}) with the help of an explicit communication protocol involving a logical paradox. On the other hand, the adopted formalism gives justice to the critique of the applicability of Hegerfeldt's theorem raised from a QFT standpoint \cite{Yngvason}. It does so by drawing attention to the characteristic scales of causality violation.

In conclusion, we put forward a new paradigm for assessing the credibility of physical theories: One needs to be able to compute the local detection statistics in an (effective) spacetime, but no details on the dynamics are prerequisite. On the theoretical side, it presents a challenge to understand the characteristic physical scales of information-theoretic-inspired post-quantum theories, such as \cite{PR_box,QuantumCausality,PawelRaviCausality,Coecke2,DAriano,BruknerX,Hardy2005}. On the practical side, it offers a powerful method for exploring the limitations of competing models of quantum wave dynamics, possibly non-unitary and/or non-linear.

\section{Acknowledgments}

\begin{acknowledgments}
The work of ME and TM was supported by the National Science Centre in Poland under the research grant Sonatina (2017/24/C/ST2/00322). PH and RH acknowledge support by the Foundation for Polish Science through IRAP project co-financed by EU within Smart Growth Operational Programme (contract no. 2018/MAB/5).
\end{acknowledgments}

\bibliography{causality_bib}

\newpage \ \newpage

\section*{Supplemental Material}

\subsection{Tightness of measures}

Any Borel probability measure $\eta$ living on a Polish space $\M$ (that is a separable completely metrizable topological space) is \emph{tight}, by which we mean that
\begin{align*}
\eta(\B) := \sup \{ \eta(\C)| \ \C \subset \B, \C \textnormal{ -- compact} \},
\end{align*}
for any measurable $\B \subset \M$ (for details, see \cite[Chapter 12]{Aliprantis}). In what follows, the term `measure' shall always stand for `Borel probability measure'. The tightness property entails, in particular, the following lemma.

\textbf{Lemma S.1.} Suppose that two measures $\eta_1, \eta_2$ on $\M$ satisfy
\begin{align*}
\eta_1(\B) < \eta_2(\B),
\end{align*}
for some measurable set $\B \subset \M$. Then there exists a \emph{compact} subset $\C \subset \B$, for which the above inequality is still valid, i.e.,
\begin{align*}
\eta_1(\C) < \eta_2(\C).
\end{align*}
Observe that $\C$ must be nonempty.

\textit{Proof}. Suppose, on the contrary, that $\eta_1(\C) \geq \eta_2(\C)$ for all compact subsets $\C \subset \B$. Then, of course, the same concerns the suprema: $\sup \{ \eta_1(\C)| \ \C \subset \B, \C \textnormal{ -- compact} \} \geq \sup \{ \eta_2(\C)| \ \C \subset \B, \C \textnormal{ -- compact} \}$. But on the strength of the tightness property, this yields $\eta_1(\B) \geq \eta_2(\B)$. \qed

\subsection{Proof of Proposition 1}

Suppose the CE condition (\ref{axiom0}) is violated, i.e., that there exist time instants $s,t$ with $s \leq t$ and a compact set $\K \subset \supp \mu_s$ such that
\begin{align*}
\mu_s(\K) > \mu_t(j^+(\K))
\end{align*}
or, switching to the more convenient notation $\mu := \mu_s$ and $\nu := \mu_t$, that
\begin{align}
\tag{$S.1$} \label{violation}
\mu(\K) > \nu(j^+(\K))
\end{align}
Our aim is to find a compact $\C \subset \cS_t \setminus j^+(\K)$ violating the NS condition (\ref{ns}). To this end, observe first that
\begin{align*}
& \nu(j^+(\K)|1) = \sum\limits_{\ss \in \{+,-,\emptyset\}} \nu(j^+(\K)|\ss,1)P(\ss|1) \\
& = 1 \cdot \mu(\K) + \nu(j^+(\K)|-,1)(1 - \mu(\K)) \geq \mu(\K)
\\
& > \nu(j^+(\K)) = \nu(j^+(\K)|0),
\end{align*}
where we have employed: the law of total probability, consistency conditions (\ref{consistency1}), Axiom 1, (\ref{violation}) and, finally, consistency condition (\ref{consistency2}). Passing to the complement, we thus have
\begin{align*}
\nu(\cS_t \setminus j^+(\K)|1) < \nu(\cS_t \setminus j^+(\K)|0).
\end{align*}
By Lemma S.1, there exists a compact subset $\C \subset \cS_t \setminus j^+(\K)$ for which the above inequality remains valid, i.e.,
\begin{align}
\tag{$S.2$} \label{violation3}
\nu(\C|1) < \nu(\C|0).
\end{align}

\subsection{Proof of Theorem 2}

The proof is somewhat technical and relies on certain notions from Lorentzian causality theory, which we briefly recall for the Reader's convenience. Namely, for any event $p \in \M$ the following sets are considered: $J^+(p), J^-(p), I^+(p), I^-(p)$ called the causal future, the causal past, the chronological future and the chronological past of $p$, respectively. Although the precise definition of these sets is not important for us here, the crucial property for the following proof is that the sets $I^\pm(p)$ are always open (topologically) and contained in $J^\pm(p)$. Moreover, we need to assume that the causal future of any compact set is topologically closed. This is true, in particular, if $\M$ is a globally hyperbolic manifold. For an excellent exposition of causality theory, the Reader is referred to \cite{MS08}.

Moving to the actual proof, let $\C \subset \cS_t \setminus j^+(\K)$ be the (nonempty) set satisfying (\ref{violation3}), and so violating the NS condition (\ref{ns}). In the first three steps of the proof we construct a compact subset $\C' \subset \C$ still violating NS, together with $q \in \M \setminus J^+(\K)$ such that $\C' \subset J^-(p)$. Then, in the last step, we show how to find $p_1,\ldots,p_k \in \M \setminus J^-(q)$ (for some $k \in \sN$) such that $\K \subset \bigcup_{i=1}^k J^+(p_i)$.

{\it Step 1.} Consider the family $\{I^-(q)\}_{q \in \M \setminus J^+(\K)}$ of open subsets of $\M$. We claim that it covers $\C$, i.e. that
\begin{align*}
\forall p \in \C \ \exists q \in \M \setminus J^+(\K) \quad p \in I^-(q).
\end{align*}
Indeed, assuming the contrary, we would have that
\begin{align*}
\exists p \in \C \ \forall q \in \M \setminus J^+(\K) \quad p \not\in I^-(q),
\end{align*}
which in fact can be equivalently written as
\begin{align*}
\exists p \in \C \quad I^+(p) \subset J^+(\K).
\end{align*}
But since $p$ lies in the closure of $I^+(p)$ and the set $J^+(\K)$ is closed \cite{MS08}, we obtain that $\C \cap J^+(\K) \neq \emptyset$, in contradiction with the inclusion $\C \subset \cS_t \setminus j^+(\K) = \cS_t \setminus J^+(\K)$.

{\it Step 2.} Since $\C$ is compact, there exists a finite subcover $\{I^-(q_i)\}_{i \in F}$, where $F$ is a finite set of indices. However, for a technical reason to become clear soon, we shall need a pairwise disjoint refinement of this subcover. To this end, one might construct the family $\U := \{U_S\}$, where the index $S$ runs over all nonempty \emph{subsets} of $F$, by defining
\begin{align*}
U_S & := \bigcap\limits_{i \in S} I^-(q_i) \setminus \bigcup\limits_{j \in F \setminus S} I^-(q_j)
\\
& = \left\{ p \in \M \ | \ \forall i \in \{1,\ldots,l\} \quad p \in I^-(q_i) \Leftrightarrow i \in S \right\}.
\end{align*}
Observe that every $U_S$ is measurable. Clearly, thus defined $\U$ is a pairwise disjoint family of sets which covers $\C$. That it is also a refinement of the cover $\{I^-(q_i)\}_{i \in F}$ stems from the fact that every $S$ is nonempty, and hence every $U_S$ is contained in at least one of the $I^-(q_i)$'s. We now claim that
\begin{align*}
\exists \, \emptyset \neq S^\ast \subset F \quad \nu(\C \cap U_{S^\ast}|1) < \nu(\C \cap U_{S^\ast}|0).
\end{align*}
Indeed, assuming that $\nu(\C \cap U_S|1) \geq \nu(\C \cap U_S|0)$ for all nonempty $S \subset F$, one would get
\begin{align*}
\nu(\C|1) & = \hspace*{-4pt} \sum\limits_{\emptyset \neq S \subset F} \hspace*{-2pt} \nu(\C \cap U_S|1) \geq \hspace*{-4pt} \sum\limits_{\emptyset \neq S \subset F} \hspace*{-2pt} \nu(\C \cap U_S|0) = \nu(\C|0),
\end{align*}
in contradiction with inequality (\ref{violation3}). It is at this step that we needed the cover to be pairwise disjoint -- otherwise we would not be able to use the measures' additivity property.

{\it Step 3.} Invoking the above Lemma, we obtain the existence of a compact $\C' \subset \C \cap U_{S^\ast}$ such that
\begin{align*}
\nu(\C'|1) < \nu(\C'|0).
\end{align*}
Moreover, picking any $i \in S^\ast$, we get $\C' \subset U_{S^\ast} \subset I^-(q_i)$ with $q_i \in \M \setminus J^+(\K)$, because only such events were involved in the construction of the original open cover in Step 2. Of course, we now set $q \cv q_i$. Since $I^-(q)$ is contained in $J^-(q)$, the first part of the proof is complete.

{\it Step 4.} Consider now the family $\{I^+(p)\}_{p \in \M \setminus J^-(q)}$. We claim  that it is an open cover of $\K$. Indeed, assuming the contrary and proceeding analogously as in Step 1., one would obtain that $\K \cap J^-(q) \neq \emptyset$, in contradiction with how $q$ was defined. By the compactness of $\K$, one can take now the finite subcover $\{I^+(p_1),\ldots,I^+(p_k)\}$. Observe that the spacetime points $p_i$ satisfy $p_i \not\preceq q$, for $i=1,\ldots,k$, as desired. Moreover, we have that $\K \subset \bigcup_{i=1}^k I^+(p_i) \subset \bigcup_{i=1}^k J^+(p_i)$, what completes the entire proof. \qed

%\pagebreak
\subsection{Proof of Theorem 3}

In what follows, $\C$ is always understood as bound by the quantifier ``for any compact $\C \subset \cS_t \setminus j^+(\K)$''.

On the strength of consistency condition (\ref{consistency2}), the NS condition (\ref{ns}) can be written as
\begin{align}
\tag{$S.3$} \label{SM_NS}
\nu(\C \, | \, 1) = \nu(\C).
\end{align}
What is more, Axiom 1 can be reexpressed as
\begin{align}
\tag{$S.4$} \label{SM_A1}
\nu( \C \, | \, + , 1) = 0.
\end{align}
Finally, with the help of one of the consistency conditions (\ref{consistency1}), we also rewrite Axiom 2 as
\begin{align}
\tag{$S.5$} \label{SM_A2}
\nu( \C \, | \, -, 1)P(-|1) = \nu(\C).
\end{align}
Invoking the law of total probability $\nu( \C \, | \, 1) = \nu( \C \, | \, +, 1)P(+|1) + \nu( \C \, | \, -, 1)P(-|1)$, one can now easily observe that any two equalities from (\ref{SM_NS},\ref{SM_A1},\ref{SM_A2}) imply the third one, what completes the first part of the proof.

It now suffices to show, e.g., that A2 implies CE. Indeed, plugging $\C := \cS_t \setminus j^+(\K)$ into (6) we obtain that
\begin{align*}
\frac{\nu(\cS_t \setminus j^+(\K))}{1 - \mu(\K)} = \nu( \cS_t \setminus j^+(\K) \, | \, - ,1) \leq 1
\end{align*}
and hence
\begin{align*}
1 - \nu(j^+(\K)) \leq 1 - \mu(\K),
\end{align*}
what is equivalent to inequality (\ref{axiom0}). This completes the proof of Theorem 3 \qed

Finally, let us demonstrate that Theorem~3 completely describes the logical relations between the four considered conditions. Namely, we shall find concrete realisations of all the logical situations not excluded by Theorem~3.

To this end, let us fix $p,q \in \cS_s$ and $p',q' \in \cS_t$ such that $p \preceq p'$, $q \preceq p'$, $q \preceq q'$ and $p \not\preceq q'$. Let us consider the following family of discrete (one- or two-point) measures
\begin{align*}
& \mu := \tfrac{1}{2} \delta_p + \tfrac{1}{2} \delta_q,
\\
& \nu := A \delta_{p'} + (1 - A) \delta_{q'},
\\
& \nu(\, \cdot \, | \, + , 1) := B \delta_{p'} + (1 - B) \delta_{q'},
\\
& \nu(\, \cdot \, | \, - , 1) := C \delta_{p'} + (1 - C) \delta_{q'},
\end{align*}
where $A,B,C \in [0,1]$ are parameters. Without much effort one can convince oneself that, for such defined measures, the four conditions considered here amount to
\begin{align*}
& \textrm{NS:} \ \ 2A = B + C, && \textrm{A1:} \ \ B = 1,
\\
& \textrm{CE:} \ \ 2A \geq 1, && \textrm{A2:} \ \ 2A = 1 + C.
\end{align*}
Now, it is not difficult to find sample values of the parameter triple $(A,B,C)$ providing a realisation of each combination of truth values assigned to the four conditions that is not excluded by Theorem 3. The following table sums that up, with T, F denoting ``true'' and ``false'', respectively.
\vspace*{5pt}
\begin{center}
 \begin{tabular}{||c|c|c|c||c||} 
 \hline
 \textrm{NS} & \textrm{A1} & \textrm{A2} & \textrm{CE} & sample $(A,B,C)$ \\
 \hline\hline
 T & T & T & T & $(1,1,1)$ \\ 
 \hline
% F & T & T & T & \Lightning \\ 
% \hline
% T & F & T & T & \Lightning \\
% \hline
 F & F & T & T & $(1,0,1)$ \\ 
 \hline
% T & T & F & T & \Lightning \\ 
% \hline
 F & T & F & T & $(1,1,0)$ \\ 
 \hline
 T & F & F & T & $(\tfrac{2}{3},\tfrac{1}{3},0)$ \\ 
 \hline
 F & F & F & T & $(1,0,0)$ \\ 
 \hline
% T & T & T & F & \Lightning \\ 
% \hline
% F & T & T & F & \Lightning \\ 
% \hline
% T & F & T & F & \Lightning \\
% \hline
% F & F & T & F & \Lightning \\ 
% \hline
% T & T & F & F & \Lightning \\ 
% \hline
 F & T & F & F & $(0,1,0), (0,1,1)$ \\ 
 \hline
 T & F & F & F & $(0,0,0)$ \\ 
 \hline
 F & F & F & F & $(0,0,1)$ \\
 \hline
\end{tabular}
\end{center}

\begin{figure}[ht]
\begin{center}
\resizebox{0.48\textwidth}{!}{\includegraphics{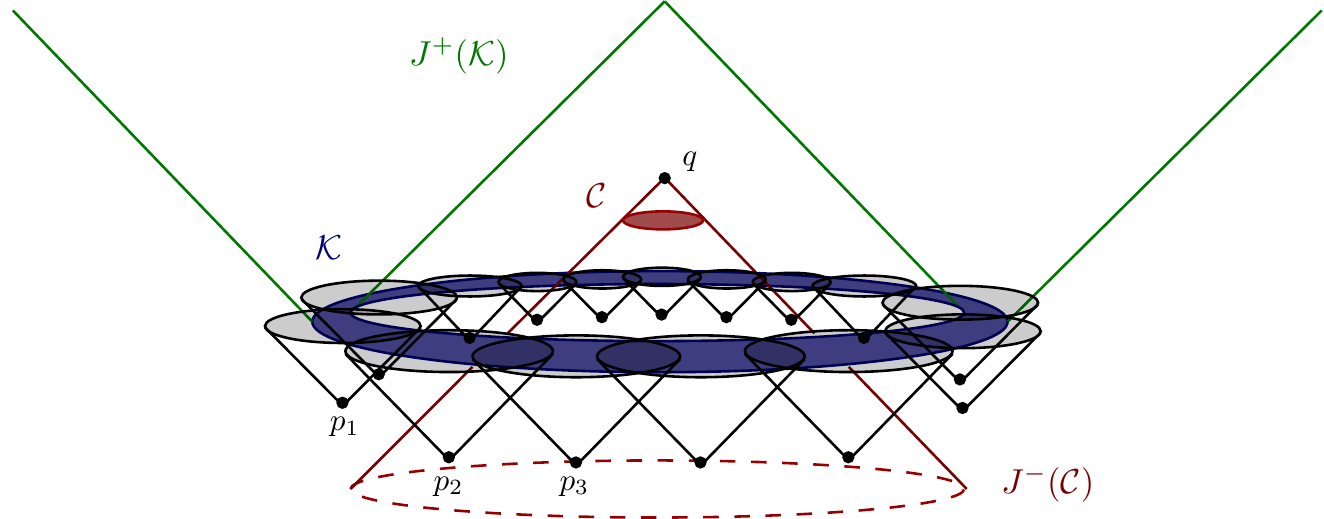}}
\caption{The figure depicts a signalling scenario in 2+1 dimensional spacetime with the set $\K$ being an annulus. To execute a superluminal communication protocol $k=15$ sending events are needed. For sake of readability, only first three events are captioned.}
\end{center}
\end{figure}

\subsection{Example of a multi-sender signalling scenario}

As explained in the main text, the shape of the set $\K$, for which the condition CE is violated, might require a coordinated measurement of multiple senders. The figure below illustrates an example when such a situation arises.

The spacetime configuration is such that there exists no single event $p$ such that $\K \subset J^+(p)$ and $\C \not\subset J^+(p)$. Consequently, although there is an information flow from $\K$ to the spacelike separated region $\C$, the communication from any $p$ with $\K \subset J^+(p)$ to any $q$ such that $\C \subset J^-(q)$ is actually subluminal.

Nevertheless, there exist events $\{p_i\}_{i=1}^k$, such that:
\begin{enumerate}
	\item $p_i \npreceq p_j$ for all $i \neq j$;
	\item $p_i \npreceq q$ for all $i$;
	\item $\K_i \vc \K \cap J^+(p_i)$ are such that $\K = \bigcup_{i=1}^{n} \K_i$.
\end{enumerate}
In this case an orchestrated action of $k$ senders is required to trigger superluminal signal towards the spacelike-separated detector in $\C$ and hence to the receiver located at $q$. Namely, we can assume that $k-1$ ``Alices'' are passive and have always-on detectors, and it is the decision of the remaining (active) sender whether to perform the measurement or not, that fixes the value of the communicated bit, in full analogy with the $k=2$ case discussed in the main text.

\end{document}